\documentclass{iau} 
\usepackage{graphicx}
\usepackage{hyperref}

\title[M~31's Last Major Merger: Constraints from the survey of PNe] 
{The Andromeda Galaxy's Last Major Merger: Constraints from the survey of Planetary Nebulae}

\author[S. Bhattacharya et al.]   
{Souradeep Bhattacharya$^1$, Magda Arnaboldi$^2$, Ortwin Gerhard$^3$, Nelson Caldwell$^4$, Chiaki Kobayashi$^5$, Francois Hammer$^6$, Yanbin Yang$^6$, Kenneth C. Freeman$^7$, Johanna Hartke$^8$, Alan McConnachie$^{9}$}

\affiliation{$^1$Inter University Centre for Astronomy and Astrophysics, Ganeshkhind, Post Bag 4, Pune 411007, India\\ {\tt email: souradeep@iucaa.in}
\\
$^2$European Southern Observatory, Karl-Schwarzschild-Str. 2, 85748 Garching, Germany
\\
$^3$Max-Planck-Institut f\"ur Extraterrestrische Physik, Giessenbachstrasse, 85748 Garching, Germany
\\
$^{4}$Harvard-Smithsonian Center for Astrophysics, 60 Garden Street, Cambridge, MA 02138, USA 
\\
$^{5}$Centre for Astrophysics Research, Department of Physics, Astronomy and Mathematics, University of Hertfordshire, Hatfield, AL10 9AB, UK 
\\
$^{6}$GEPI, Observatoire de Paris, Universit\'e PSL, CNRS, Place Jules Janssen, F-92195 Meudon, France 
\\
$^{7}$Research School of Astronomy and Astrophysics, Mount Stromlo Observatory, Cotter Road, ACT 2611 Weston Creek, Australia
\\
$^{8}$Finnish Centre for Astronomy with ESO (FINCA), University of Turku, FI-20014 Turku, Finland
\\
$^{9}$NRC Herzberg Institute of Astrophysics, 5071 West Saanich Road, Victoria, BC V9E 2E7, Canada
}

\pubyear{2023}
\volume{377}  
\jname{Early Disc-Galaxy Formation: From JWST to the Milky Way}
\editors{F. Tabatabaei \& B. Barbuy, eds.}
\begin{document}

\maketitle

\begin{abstract}
The Andromeda galaxy (M~31) has experienced a tumultuous merger history as evidenced by the many substructures present in its inner halo. We use planetary nebulae (PNe) as chemodynamic tracers to shed light on the recent merger history of M~31. We identify the older dynamically hotter thicker disc in M~31 and a distinct younger dynamically colder thin disc. The two discs are also chemically distinct with the PN chemodynamics implying their formation in a `wet' major merger (mass ratio $\sim$ 1:5) $\sim$2.5--4~Gyr ago. From comparison of PN line-of-sight velocities in the inner halo substructures with predictions of a major-merger model in M~31, we find that the same merger event that formed the M~31 thick and thin disc is also responsible for forming these substructures. We thereby obtain constraints on the recent formation history of M~31 and the properties of its cannibalized satellite.

\end{abstract}

\firstsection 
\section{Introduction}

Imaging surveys of M~31 covering almost 100 sq. deg. on the sky (e.g. PAndAS; McConnachie et al. \cite{Mcc18}) measuring the projected number density of resolved red giant branch (RGB) stars have revealed several faint and diffuse substructures in the M~31 halo. Chemodynamic measurements of M~31's stellar population is required for shedding light on its tumultuous recent merger history. Owing to its large angular span on the sky and limitations of present day instrumentation, chemodynamic measurements from absorption lines features of individual RGB stars over the entirety of the M~31 disc and inner halo is challenging (E.g. SPLASH -- Guhathakurta et al. \cite{G04}; DESI -- Dey et al. \cite{Dey23}), with substantial contamination from MW halo stars. 

Planetary Nebulae (PNe), emission-line nebulae in the late stages of stellar evolution for stars with initial masses $\sim$0.8--8 M$\rm_\odot$, are discrete tracers of galaxy stellar population properties such as light, kinematics and chemistry (e.g. Hartke et al. \cite{H22}). At the distance of M~31, PNe can be identified over the entire span of its disc and inner-halo in reasonable time with wide-field imaging and spectroscopic follow-up of brighter emission line features in PNe efficiently allows for chemodynamic measurements. Additionally the MW halo is transparent to PNe (any MW halo PNe would be much brighter and easily identifiable) and background galaxy contamination is minimal at the distance of M~31. We thus utilise PN chemodynamics to understand the recent formation history of M~31.

\section{The survey of Planetary Nebulae in M~31}

\begin{figure}
	\centering
	\includegraphics[height=0.8\columnwidth,angle=0]{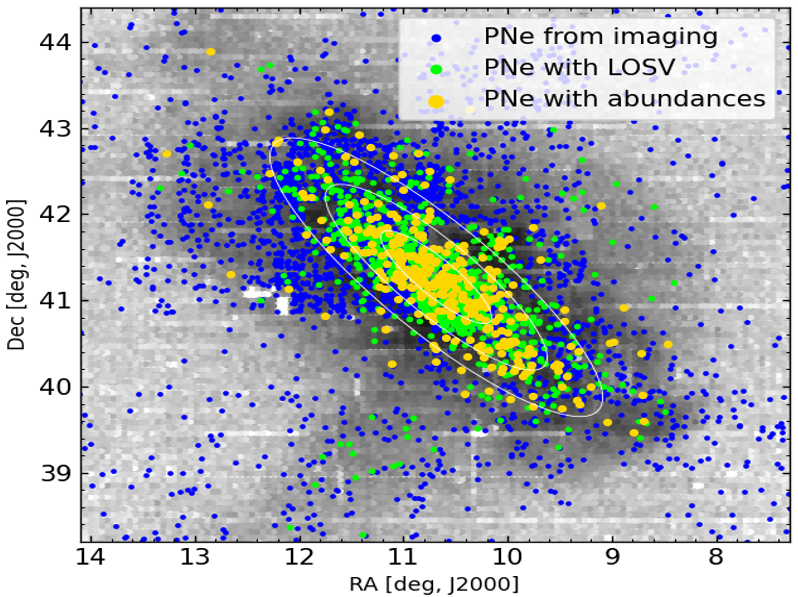}
	\caption{The number density map of RGB stars from PAndAS (McConnachie et al. \cite{Mcc18}), binned for visual clarity, is shown in grey. The PNe candidates identified in the imaging survey (within this field of view) are marked in blue (see Bhattacharya et al. \cite{Bh21} for the full sample); those with LOSV measurements are marked in green;  those with chemical abundance measurements are marked in yellow (Bhattacharya et al. \cite{Bh22}).}
	\label{fig:spat}
\end{figure}

PN candidates were photometrically identified in a 54 sq. deg. [OIII] 5007 \AA\ narrow-band and broad g-band imaging survey (Bhattacharya et al. \cite{Bh19a,Bh21}) covering the disc and inner halo of M31 with MegaCam at the CFHT. We identified 5265 PN candidates in M~31, the largest PN sample in any galaxy, of which 4085 were newly discovered (Bhattacharya et al. \cite{Bh21}). Figure~\ref{fig:spat} shows a part of the sample of the identified PNe.

Spectroscopic follow-up of a complete sub-sample of these PN candidates were carried out with the Hectospec MOS spectrograph (covering $\sim$3685--9200~\AA) on the MMT Telescope (Fabricant et al. \cite{fab05}). The sample of 413 PN spectra (including those found by Sanders et al. \cite{san12}) with confirmed detection of the [OIII] 4959/5007~\AA~ emission lines, allowing us to obtain their LOSV measurements, were first utilised for the kinematic analysis of the M~31 disc presented in Bhattacharya et al. (\cite{Bh19b}). The survey was extended in Bhattacharya et al. (\cite{Bh22}) to include a total of 1251 PN LOSV measurements in M~31. Their spatial distribution is shown in Figure~\ref{fig:spat}. Within the M~31 disc, the temperature-sensitive [OIII] 4363~\AA~line was detected in a magnitude limited sample of 205 PNe (see Figure~\ref{fig:spat}), enabling the computation of oxygen and argon abundances. The details of the chemical abundance analysis are presented in Bhattacharya et al. (\cite{Bh22}).

\section{The kinematically and chemically distinct thin and thicker discs of M~31}

In Bhattacharya et al. (\cite{Bh19b}), through velocity dispersion profiles of high- and low- extinction PNe in the M~31 disc, we respectively identify the kinematically distinct dynamically colder thin and hotter thicker disc. At the equivalent radius of the solar neighbourhood in M~31, the thin and thicker disc of M~31 are respectively about twice ($\sigma_{\phi}\sim61$~km~s$^{-1}$) and thrice ($\sigma_{\phi}\sim101$~km~s$^{-1}$) as dynamically hot as the MW thin disc population of the same age (see also Dorman et al. \cite{dor15}). The age-velocity dispersion relation in the M~31 disc is consistent with a $\sim$1:5 major merger in M~31 $\sim$2.5--4.5 Gyr ago. A minor merger ($\sim$~1:20) would not be able to dynamically heat the M~31 disc to the measured thicker disc velocity dispersion. See Bhattacharya et al. (\cite{Bh19b}) for details.

In Bhattacharya et al. (\cite{Bh22}), we found that the kinematically distinct discs are also chemically distinct. The thicker disc has near-flat radial abundance gradients in oxygen ($0.006 \pm 0.003$ dex/kpc) and argon ($-0.005 \pm 0.003$ dex/kpc), typical of discs that have experienced a recent major merger, while the thin disc had steeper gradients for both oxygen ($-0.013 \pm 0.006$ dex/kpc) and argon ($-0.018 \pm 0.006$ dex/kpc) consistent with that of the MW thin disc (accounting for different disc scale lengths). M31 PN abundance gradients are discussed in detail in Bhattacharya et al. (\cite{Bh22}). 

In Arnaboldi et al. (\cite{Ar22}), we identified that the [Ar/H] vs [O/Ar] plane for emission line nebulae is analogous to the [Fe/H] vs [$\alpha$/Fe] plane for stars, and exploration of the M~31 PN population in this plane allowed us to constrain the chemical enrichment and star formation history of the thin and thicker disc in M~31. Through the use of galactic chemical evolution models (Kobayashi et al. \cite{K20}), we found that the older M~31 thicker disc PNe show monotonic chemical enrichment in an extended star formation episode up to super-solar values, while the younger M~31 thin disc is formed $\sim$2--4~Gyr ago from accreted metal-poor gas that mixed with the relatively metal-rich M~31 ISM (consistent with that observed from the star-formation history of resolved stars in the M~31 disc; Williams et al. \cite{wil17}). This has been presented in detail in Arnaboldi et al. (\cite{Ar22}).

\section{The origin of M~31 inner-halo substructures}

In Bhattacharya et al. (\cite{Bh21}), we investigated the properties of the PN Luminosity Function (PNLF) for the stellar populations in the M~31 disc and inner-halo substructures (namely Giant Stream, NE- and W-shelves and G1-clump). We found that the GS and NE-Shelf are consistent with being composed of stellar debris from an in-falling satellite, while the G1-Clump is linked to the pre-merger M~31 disc (see also Faria et al. \cite{F07}).

In Bhattacharya et al. (\cite{Bh23}), we present the first LOSV measurements of PNe co-spatial with the aforementioned M~31 inner-halo substructures. We compare the measured LOSV distribution and observed projected radial distance vs. LOSV phase space of the PNe (and RGBs where available) to their simulated analogues in the major-merger (mass-ratio $\sim$1:4) simulation by Hammer et al. (\cite{ham18}). The NE- and W-shelves, and Giant Stream are consistent with being satellite debris (with some fraction of host material) while the G1-clump is consistent with being perturbed M~31 disc material. We thus link the formation of the aforementioned substructures in a single unique event, consistent with a major merger. In particular, the G1-clump cannot be formed from perturbed disc material in a minor merger. Details are provided in Bhattacharya et al. (\cite{Bh23}).
\section{Constraints on the cannibalized satellite properties}

From the aforementioned results, we find that M~31 experienced a 'wet' major merger (mass ratio $\sim$ 1:5) $\sim$2.5--4 Gyr ago (Bhattacharya et al. \cite{Bh23}). This event formed the M~31 thicker disc and the G1-clump from the pre-merger M~31 disc stars, while the thin disc was formed from gas brought in by the satellite mixed with the pre-enriched gas in the pre-merger disc. The satellite debris material formed the other inner-halo substructures. 

We thus obtain important constraints on the properties of the now cannibalized satellite. It would have been gas-rich ($\sim$20\% gas fraction; Arnaboldi et al. \cite{Ar22}) with a mass about twice that of M33 and would have fallen into M~31 along the Giant Stream. Further constraints would be forthcoming from improved chemodynamical measurements of RGBs and PNe, and updated major merger simulations.


\begin{acknowledgements}
SB thanks the organizers of the IAU Symposium 377 for the opportunity to present a contributed talk. Based on observations obtained with MegaPrime/ MegaCam at the CFHT. Based on observations obtained at the MMT Observatory. SB is funded by the INSPIRE Faculty award (DST/INSPIRE/04/2020/002224), Department of Science and Technology (DST), Government of India. SB acknowledges support from the IMPRS on Astrophysics at the LMU Munich during his PhD. A preliminary version of this work appears in his PhD thesis (\cite{Bh20}).   
\end{acknowledgements}

{}

\end{document}